\documentclass{IEEEtran}
\usepackage{cite}
\usepackage{amsmath,amssymb,amsfonts}
\usepackage{graphicx}
\usepackage{textcomp,nicefrac}
\usepackage[OT1]{fontenc} 
\usepackage{makecell} 
\def\BibTeX{{\rm B\kern-.05em{\sc i\kern-.025em b}\kern-.08em
T\kern-.1667em\lower.7ex\hbox{E}\kern-.125emX}}
\begin{document}
\title{A High Spatial Resolution Muon Tomography Prototype System based on Micromegas Detector}
\author{Yu Wang, Zhiyong Zhang, Shubin Liu*, Zhongtao Shen, Changqing Feng, Jianguo Liu and Yulin Liu
\thanks{This study was supported by The National Science Fund for Distinguished Young Scholars (Grant No.12025504). (Corresponding Author Shubin Liu).}
\thanks{Yu Wang, Zhiyong Zhang, Shubin Liu, Zhongtao Shen, Changqing Feng, Jianguo Liu and Yulin Liu are with State Key Laboratory of Particle Detection and Electronics, University of Science and Technology of China, No.96, Jinzhai Road, Hefei 230026, Anhui, China and also with Department of Modern Physics,  University of Science and Technology of China, No.96, Jinzhai Road, Hefei 230026, Anhui, China (e-mail: liushb@ustc.edu.cn ).}}

\maketitle

\begin{abstract}
Cosmic ray muon has strong penetrating power and no ionizing radiation hazards, which makes it an ideal probe for detecting special nuclear materials. In this paper, a high spatial resolution muon tomography system based on Micromegas detectors is proposed to optimize the imaging time and quality. The proposed system includes eight  Micromegas detectors based on the thermal bonding technique and a scalable readout system. In addition, a multiplexing method based on position encoding is developed to reduce the number of electronics channels by order of magnitude. The spatial resolution of the proposed system with encoding readout can reach a value of hundred micrometers. Finally, a tomography test is performed, and test results show that this proposed system can image 2-cm objects and distinguish different materials.

\end{abstract}


\section{Introduction}
\label{sec:introduction}
\begin{figure}[htbp]
\centerline{\includegraphics[width=3.5in]{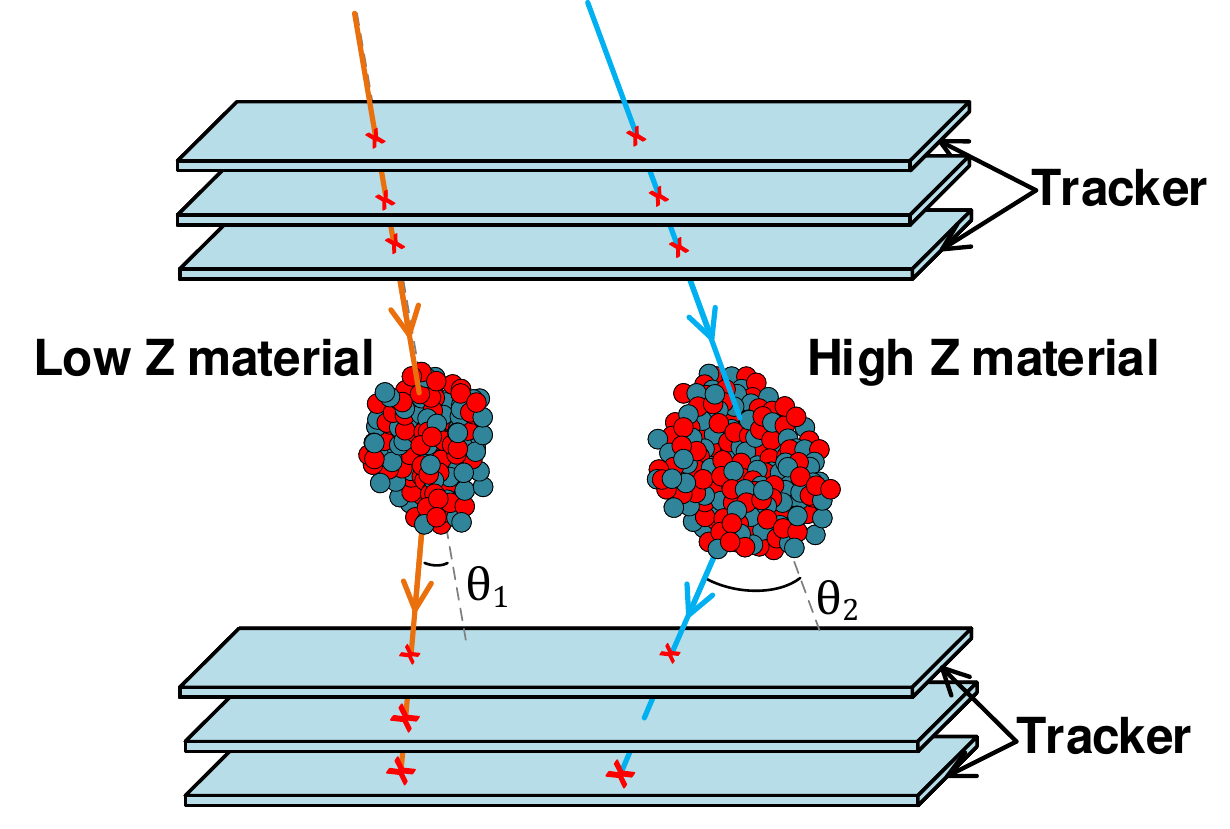}}
\caption{Principle of muon tomography. The material in the imaging area changes the direction of the muon by multiple Coulomb scatters. The scatter angle and scatter density can be used to reconstruct an object. }
\label{fig_1}
\end{figure}

\IEEEPARstart{R}{adiation} imaging technology has been widely used in the field of interior structure detection. However, detecting special nuclear materials (SNMs) can be challenging when objects are well wrapped by nuclear shielding materials. A potential method for addressing this challenge is to use cosmic ray muons to detect the inner structure of these materials, known as muon scattering tomography (MST)~\cite{bib:bib1}. Cosmic ray muons can penetrate through nuclear shielding material due to its low energy loss and long mean lifetime. In the MST, the primary interaction between cosmic muons and materials is multiple Coulomb scattering. When muons are passing through materials, their trajectories will be perturbed by the nucleus, and the scattering angle and scattering density are different for various materials. As shown in \figurename~\ref{fig_1}, to measure the scattering process  of cosmic ray muons, several tracker detectors are placed above and below the objects to record the incident and scattering trajectories. The scattering angle is related to the properties of the material under test. The projected angle can approximate a Gaussian distribution with a mean value of zero and the root-mean-square (RMS) width given by (\ref{eq1}), where $\beta \mathrm{c}$ and $p$ denote the velocity and momentum of muon, respectively, and $\mathrm{x/X_0}$ is the thickness of the material in radiation length~\cite{bib:bib2}. As shown in \figurename~\ref{fig_1}, when a muon is passing through materials with a high  atomic number, a large deflection angle will be generated.

\begin{equation}
\label{eq1}
\sigma\left(\theta_{0}\right)=\frac{13.6 \mathrm{MeV}}{\beta c p} \sqrt{x / X_{0}}\left[1+0.038 \ln \left(x / X_{0}\right)\right]
\end{equation}

In 2003, Los Alamos Laboratory first developed the muon tomography technique based on multiple scattering \cite{bib:bib3}. Five hundred seventy-six drift tubes were divided into two groups to record the incident and emergent trajectories. Using the algorithm named the Point of Closest Approach (PoCA), a tungsten cylinder was reconstructed correctly within the imaging area~\cite{bib:bib4}. In recent years, a variety of detectors combined with novel readout methods were applied to the muon tomography, for instance, drift chambers~\cite{bib:bib5}\cite{bib:bib6},  plastic scintillators~\cite{bib:bib7,bib:bib8,bib:bib9}, RPCs (Resistive Plate Chambers) \cite{bib:bib10}, MRPCs (Multigap Resistive Plate Chambers)~\cite{bib:bib11}, and GEMs (Gas Electron Multipliers)~\cite{bib:bib12}. These methods have certain advantages and disadvantages in terms of the detection area, spatial resolution, acceptance, robustness, and miscellaneous applications.

Using finer resolution detectors can provide more accurate results and make the facility more compact. The spatial resolution and detection area are two key parameters to optimize the imaging time and accuracy. The imaging time is determined by the number of muon events  that penetrate the sample material and detectors. A muography facility with a larger detector area has a greater acceptance and can record muon from more directions, which can keep the experimental time to a reasonable range. Finer resolution detectors can provide a more accurate result and keep the facility more compact. A study with Geant4 simulation has shown that the resolution of an MST system with detectors spaced at 5 mm should be  better than $200~\mu m$~\cite{bib:bib13}.

With the development of nuclear instruments, micro-pattern gaseous detectors have become a feasible option for the MST facility, which can achieve a resolution of a few hundred micrometers at a reasonable cost. In particular, the Micromegas detector can achieve a resolution of less than  $100~\mu m$ with thousands of $cm^2$ active areas~\cite{bib:bib_ALTAS_MM}. A large detection area with fine resolution increases the complexity of the readout system design. However, with the proper multiplexing method, a balance  between detection area and readout complexity can be achieved.

In this paper, a high spatial resolution tomography prototype based on Micromegas detectors is developed. To achieve high resolution, an MST system requires using thousands of electronics channels, which can cause certain issues regarding system complexity, power consumption,  and cooling. Due to the low flux and sparse hit of cosmic ray muons, a multiplexing readout method is introduced in this study, and the number of channels required by the readout electronics is reduced by one order of magnitude. In addition, verified front-end electronics cards (FEC) are used, and a general data acquisition (DAQ) board adapted to different scales of imaging experiments is designed. Finally, tomography experiments are performed on objects with a size of several centimeters .

\section{System Design and Experimental Setup}
A tomography prototype was built to verify the imaging capability of Micromegas detectors and to study the readout scheme of muon tomography. The prototype included Micromegas detectors, encoding readout circuits, FECs, and DAQ board. The system design and specific parameters of each part are described below.

\subsection{Resistive Micromegas Detector}
\begin{figure}[htbp]
\centerline{\includegraphics[width=3.5in]{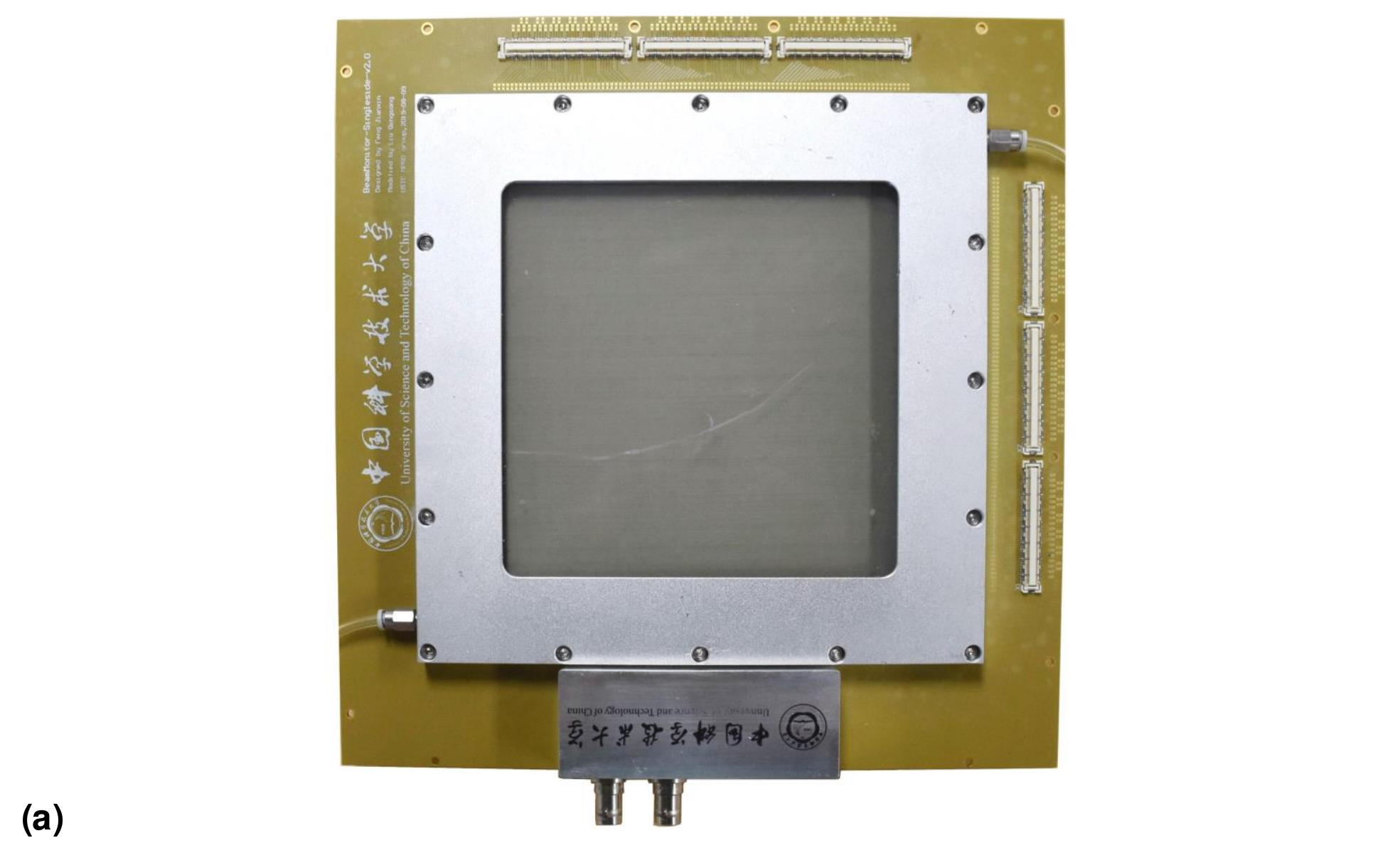}}
\centerline{\includegraphics[width=3.5in]{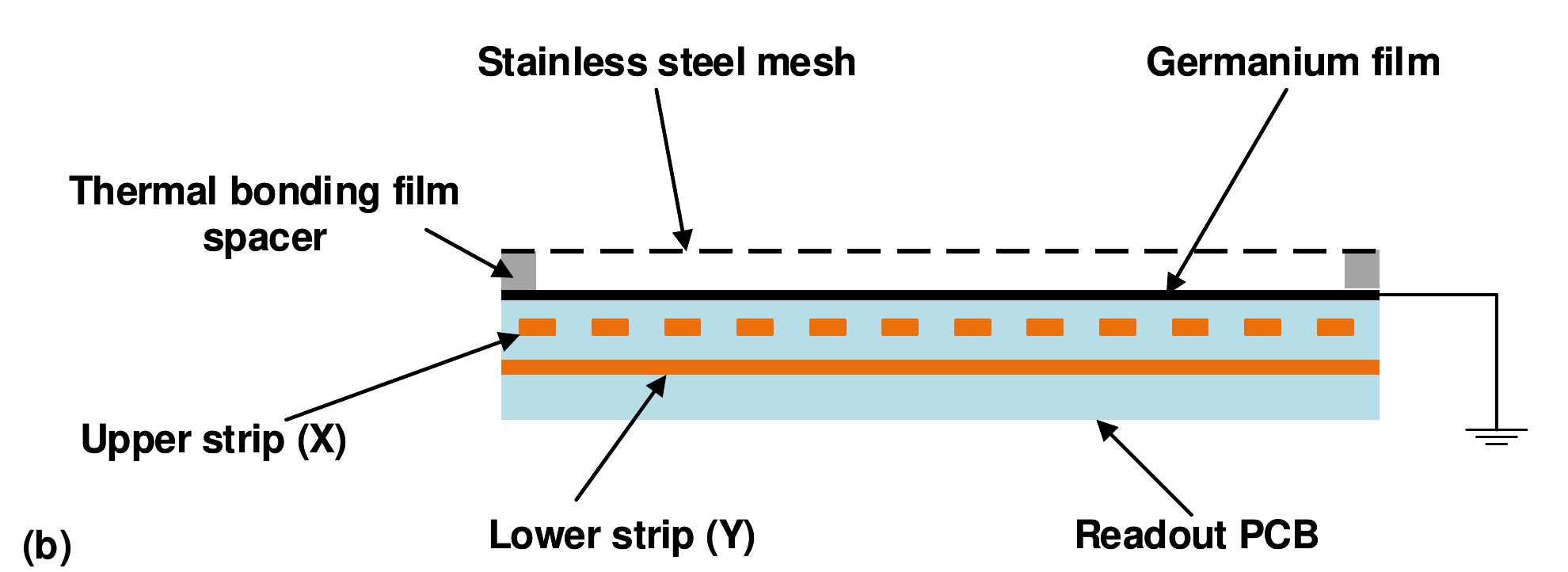}}
\caption{(a) Photograph of a Micromegas detector used for muon tomography; (b) side view of the Micromegas structure with 2D readout strips.}
\label{fig_2}
\end{figure}

The Micromegas detectors were manufactured by the thermal bonding method (TBM), and the sensitive area was $150~ mm \times 150~mm$, as shown in \figurename~\ref{fig_2}(a) [14]. This manufacturing method is a simple and efficient way to manufacture Micromegas detectors while maintaining the good performance of the detectors. Two orthogonal readout strips were placed into separate inner layers of the readout PCB (Printed Circuit Board), and the pitch was $400 ~\mu m$. The width of the strips of the x-dimension in the upper layer was $80 ~\mu m$, and the width of the strips of the y-dimension in the lower layer was $320~ \mu m$, as shown in \figurename~\ref{fig_2}(b). A germanium layer was coated on the surface of the PCB and used as a resistive anode. The drift gap between the drift cathode and the mesh was 5 mm, and the avalanche gap was about $110~ \mu m$. A gas mixture of argon and $\mathrm{CO_2}$ (7\%) was used as an active medium, and the working voltages of the mesh and drift cathode were -540~V and -720~V, respectively.

The performances of the detectors were thoroughly tested with the radiation source and electron beams~\cite{bib:bib14}. Using the $^{55}\mathrm{Fe}$ radioactive source (5.9 keV X-ray), an energy resolution of 16\% and a gain of over $10^4$ were achieved. Furthermore, a spatial resolution of $\mathrm{65 ~\mu m}$ and an efficiency of 98\% were achieved using the 5-GeV electron beam at DESY (Deutsches Elektronen–Synchrotron). These performances indicated that the Micromegas detectors with TBM could satisfy the requirement of muon tomography.

\subsection{Position Encoding Readout Method} 
\label{sub:position_encoding_readout_method}
To compress readout channels, a channel multiplexing method with the position encoding scheme was developed and implemented. Each detector strip could be mapped only to one readout channel, whereas each electronics channel could read out multiple detector strips with specific rules. In the encoding scheme, two adjacent detector strips should be mapped to a unique pair of unordered readout channels. The hit position should always be continuous at the detector side so that the result of the inferred hit strips would be continuous.

\begin{figure}[htbp]
	\centerline{\includegraphics[width=3.5in]{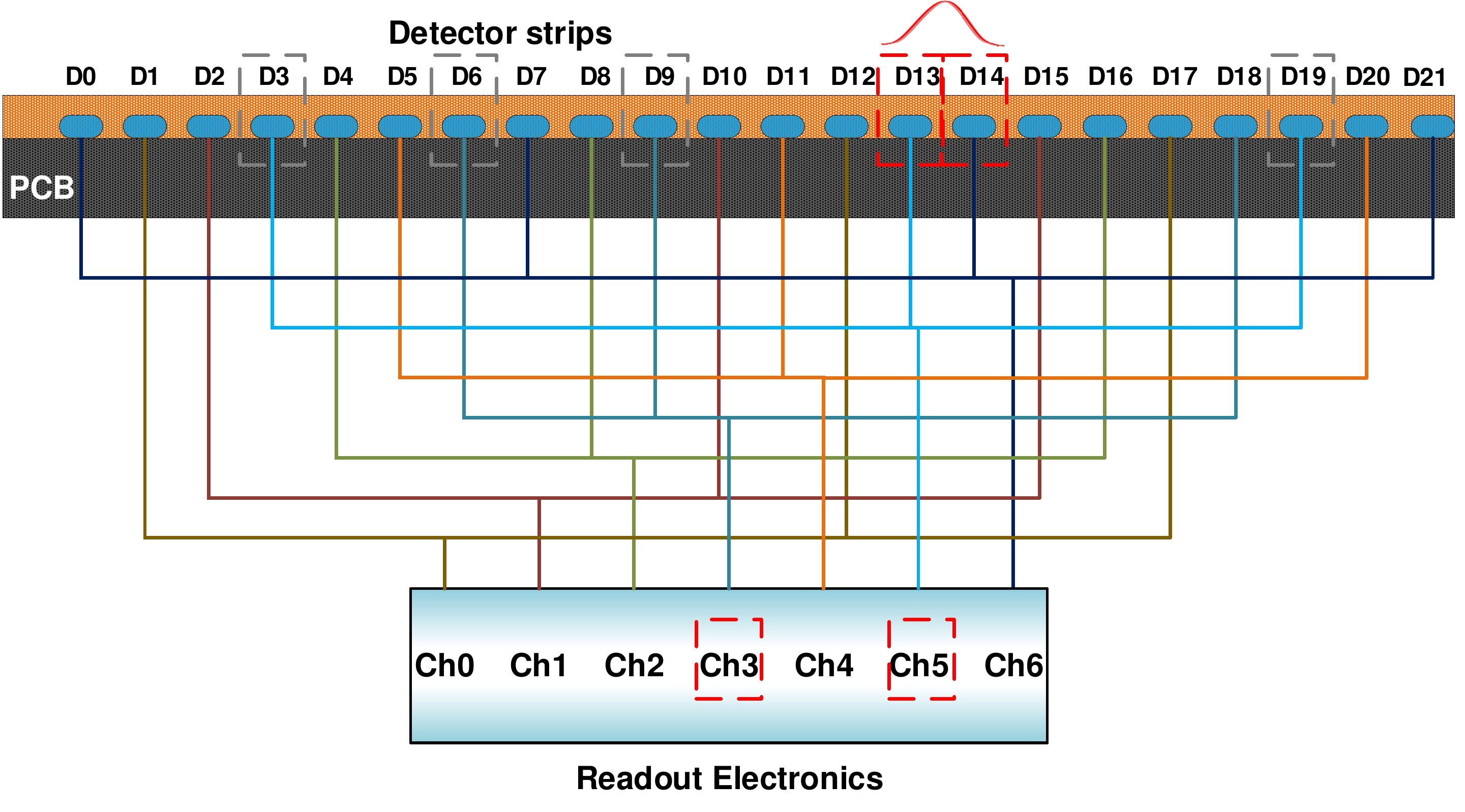}}
\caption{ An example of encoding 22 detector strips to seven readout channels.}
\label{fig_3}
\end{figure}

Generally, assume that two neighboring detector strips $i$ and $(i + 1)$ are read out by two readout channels $a$ and $b$, respectively, and the two readout channels are also connected to the other non-neighboring detector strips. If an ionization signal is recorded only by channels $a$ and $b$, the hit strips will be $i$ and $(i + 1)$. Therefore, if any continuous combination of detector strips connects to a unique pair of readout channels, the hit position of the signal can be precisely derived. \figurename~\ref{fig_3} shows an example of an encoding scheme of 22 detector strips to seven readout electronics; where Ch0-Ch6 denote the electronics channels, and D0–D21 denote the detector strips. For instance, if an ionization happens at two neighboring strips D13 and D14, the corresponding signals will be recorded by readout channels Ch3 and Ch5. In turn, if Ch3 and Ch5 are fired, the possible hit detector channels may be D3, D6, D9, D13, D14, and D19. Among these possibilities, the only continuous combination is D13 and D14. The above-mentioned encoding relationship is shown in \tablename~\ref{tab1}, and it can be divided into three encoding groups.

\begin{table}[htbp]
\centering
\caption{Encoding 21 detector strips to 7 readout channels}
\label{tab1}
\setlength{\tabcolsep}{3pt}
\begin{tabular}{m{40pt}<{\centering}|m{15pt}<{\centering} m{15pt}<{\centering} p{15pt}<{\centering} p{15pt}<{\centering} p{15pt}<{\centering} p{15pt}<{\centering} p{15pt}<{\centering} p{15pt}<{\centering}}
\hline
Detector channel &  D0 & D1 & D2 & D3 & D4 & D5 & D6 \\

Readout channel & Ch6 & Ch0 & Ch1 & Ch5 & Ch2 & Ch4 & Ch3 \\
\hline
Detector channel &  D7 & D8 & D9 & D10 & D11 & D12 & D13 \\

Readout channel & Ch6 & Ch2 & Ch3 & Ch1 & Ch4 & Ch0 & Ch5 \\
\hline
Detector channel &  D14 & D15 & D16 & D17 & D18 & D19 & D20 & D21 \\

Readout channel & Ch6 & Ch1 & Ch2 & Ch0 & Ch3 & Ch5 & Ch4 & Ch6 \\
\hline
\end{tabular}
\end{table}

\begin{figure}[htbp]
\centerline{\includegraphics[width=2.4in]{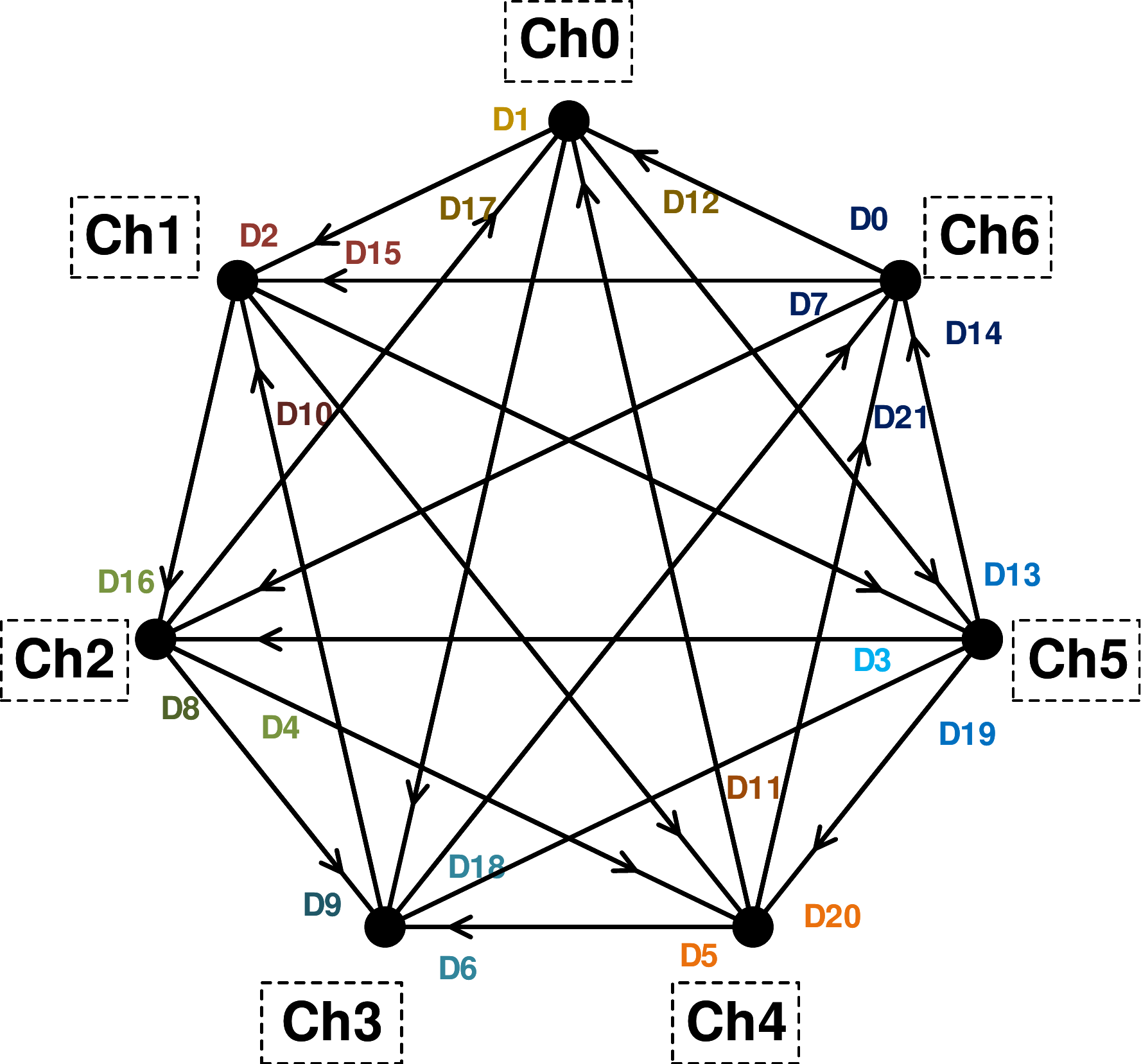}}
\caption{ The encoding graph; vertices Ch0–Ch6 stand for readout electronics channels and edges D0–D21 stand for connected detector strips.}
\label{fig_4}
\end{figure}

The encoding relationship can be simplified to the Eulerian circuit problem of a complete graph, where vertices correspond to readout channels, and edges correspond to doublet combinations of continuous detector strips~\cite{bib:bib15}. In the Eulerian circuits, each edge is passed exactly once. Therefore, a combination of adjacent detector strips is connected to a unique pair of readout channels only once. \figurename ~\ref{fig_4} shows the construction process of the Eulerian circuit consisting of seven readout channels that correspond to the sequence of $\mathrm{Ch} 6 \rightarrow \mathrm{Ch} 0 \rightarrow \mathrm{Ch} 1 \rightarrow \mathrm{Ch} 5 \rightarrow \mathrm{Ch} 2 \rightarrow \mathrm{Ch} 4 \rightarrow \mathrm{Ch} 3 \rightarrow \mathrm{Ch} 6 \rightarrow \mathrm{Ch} 2 \rightarrow \mathrm{Ch} 3 \rightarrow \mathrm{Ch} 1 \rightarrow \mathrm{Ch} 4 \rightarrow \mathrm{Ch} 0 \rightarrow \mathrm{Ch} 5 \rightarrow \mathrm{Ch} 6 \rightarrow \mathrm{Ch} 1 \rightarrow \mathrm{Ch} 2\rightarrow \mathrm{Ch} 0\rightarrow \mathrm{Ch} 3 \rightarrow \mathrm{Ch} 5 \rightarrow \mathrm{Ch} 4 \rightarrow \mathrm{Ch} 6$; this sequence is determined based on the connection relationship of D0–D21, and a general rule is described in the following part. In the encoding graph, each vertex is connected with more than one edge so that each readout channel can be multiplexed to more than one detector strip.

A well-designed encoding scheme should make an interval between the same multiplex channels as large as possible. If and only if the interval is larger than the total number of hit strips, the decoding result will be an actual result. 
The problem is described as the Eulerian recurrent length  (ERL) of a complete graph, and it has been proven that the minimum length is $(n-4)$, where n is the number of vertices~\cite{bib:bib16}\cite{bib:bib17}. 
The Eulerian path of a complete graph was constructed in~\cite{bib:bib17},  where n was an odd number; a complete graph is an Eulerian graph, if and only if the number of vertices is odd. In this study, the Eulerian path with an even number of vertices was constructed, in which the minimum ERL was still $(n - 4)$. The construction method is defined by:
\begin{equation*}
C_{n}:=\left\{\begin{array}{l}
H_{0} \rightarrow H_{2} \rightarrow \cdots \rightarrow H_{(n-6) / 2} \rightarrow H_{(n-4) / 2} \rightarrow \\
H_{(n-4) / 2-2} \rightarrow \cdots \rightarrow H_{1} \rightarrow n-2, \text { if } n \bmod 4=2 \\
H_{0} \rightarrow H_{2} \rightarrow \cdots \rightarrow H_{(n-4) / 2} \rightarrow H_{(n-6) / 2} \rightarrow \\
H_{(n-6) / 2-2} \rightarrow \cdots \rightarrow H_{1} \rightarrow n-2, \text { if } n \bmod 4=0
\end{array}\right.
\end{equation*}
where  $H_{k}=n-2 \rightarrow v_{0}(k) \rightarrow v_{1}(k) \rightarrow \cdots v_{(n-4) / 2}(k) \rightarrow n-1 \rightarrow v_{(n-2) / 2}(k) \rightarrow \cdots \rightarrow v_{n-3}(k)$ , and $v_i (k)\quad (i \leq n-3)$ is defined as follows:
\begin{equation*}
v_{i}(k)=\left\{\begin{array}{l}
k, \quad i=0 \\
\left(v_{i-1}(k)+i\right) \bmod (n-2), i>0 \text { and } i \bmod 2=1 \\
\left(v_{i-1}(k)-i\right) \bmod (n-2), \text { otherwise }
\end{array}\right.
\end{equation*}

In the Eulerian path, the readout electronics channels were marked as $0,1, \ldots,(\mathrm{n}-1)$ and each $v_i (k)$ in $C_n$ was regarded as a detector strip. The mapping scheme was one-to-one matching between the readout channels and detector strips within any $(n - 4)$ continuous detector strips.

\begin{figure}[htbp]
\centerline{\includegraphics[width=3.5in]{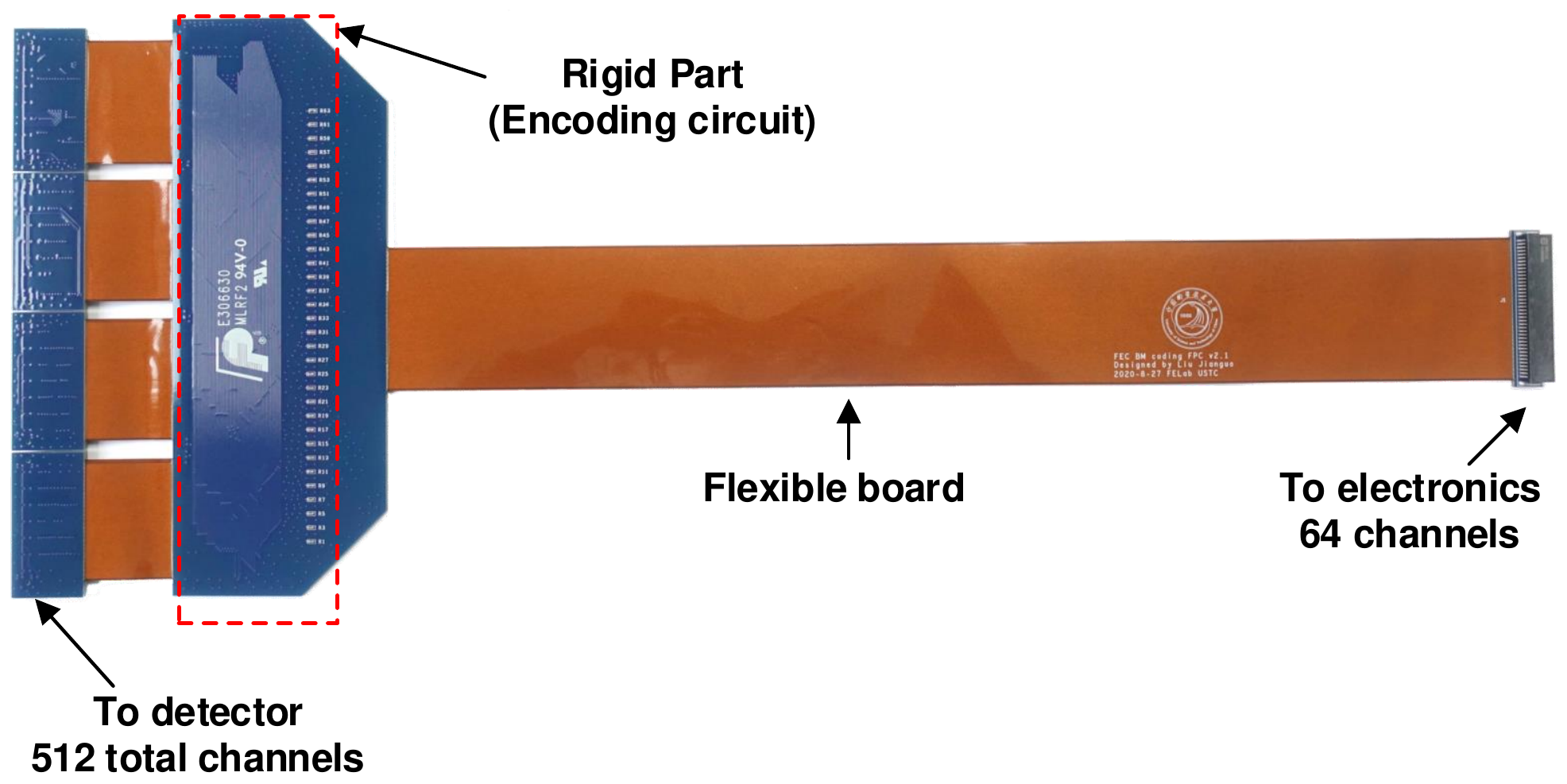}}
\caption{Photograph of the encoding readout connector.}
\label{fig_5}
\end{figure}

Using the position encoding algorithm, an encoding scheme with 64 readout electronics channels was constructed, and the theoretical number of detector channels was 1,986. The encoding table was truncated to accommodate the detector and the readout electronics channels, and an encoding board of 512 detector strips was designed for 64 electronics channels. \figurename~\ref{fig_5} shows the photo of the encoding readout flexible board made of Kapton. Each encoding board could readout one dimension of a detector.

\subsection{FEC} 
\label{sub:front_end_electronics_card}
When muons hit the detector, the induced charges will spread over several readout strips due to the diffusion \cite{bib:bib_diffusion} . The charge centroid method can improve the achievable spatial resolution, which uses the charge-over-position distribution to retrieve the hit position~\cite{bib:bib_diffusion,bib:bib_CCM,bib:bib_CCM1}. Therefore, a sufficient charge sensitivity readout system was established to acquire the charge information and hit strips. The FEC design based on the AGET (Asic for General Electronics for Tpc)~\cite{bib:bib_AGET} and verified by the PandaX III (Particle AND Astrophysical Xenon experiment III) project~\cite{bib:bib_FEC} was adopted. The AGET is an analog readout chip that contains 64 channels, each of which consists of a charge-sensitive preamplifier (CSA), an analog shaper, a discriminator for a trigger, and a 512-SCA (Switched Capacitor Array). The sampled signal will be driven to a pair of differential pins under the control of readout circuits. 

\begin{figure}[htbp]
\centerline{\includegraphics[width=3.5in]{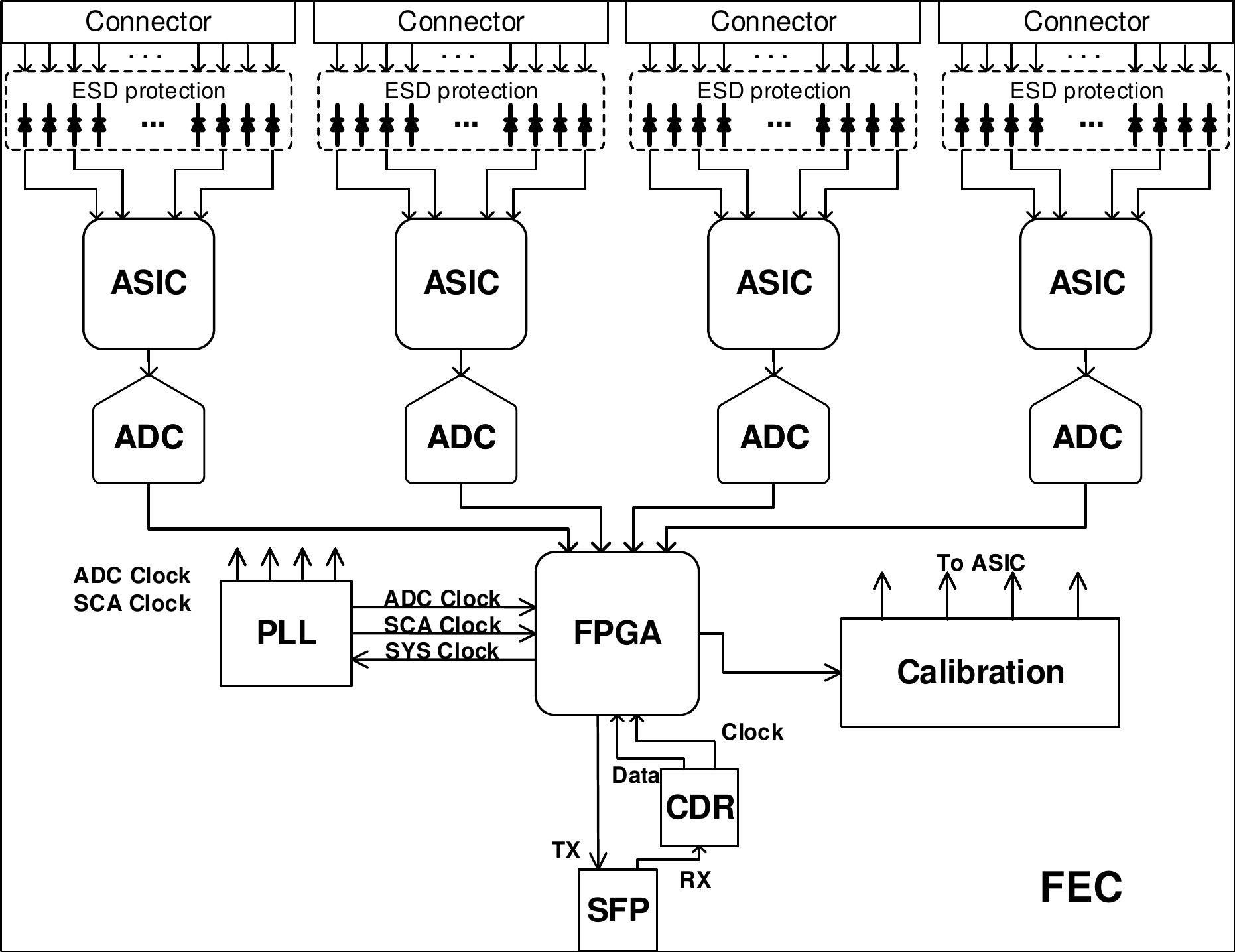}}
\caption{Schematic diagram of the FEC.}
\label{fig_6}
\end{figure}

As shown in \figurename~\ref{fig_6}, a single FEC used four AGET chips to sample the encoded signals. The output of each AGET chip was digitalized by a single-channel 12-bit analog-to-digital converter (ADC). Then, the digitalized data were readout, reconstructed, and transmitted by a field-programmable gate array (FPGA). When receiving the trigger signal, the FPGA would stop the acquisition phase of the AGET after a programmable delay and start the ADC conversation process. Due to the low trigger rates, a channel-to-channel threshold was set to suppress the channels without signal. All the digitization data whose values were greater than the preset thresholds, the timestamp, trigger number, and checksum would be constructed into a user-defined data frame and transmitted to the back-end DAQ board.

\subsection{Data Acquisition Board} 
\label{sub:data_acquisition_board}
\begin{figure}[htbp]
\centerline{\includegraphics[width=3.5in]{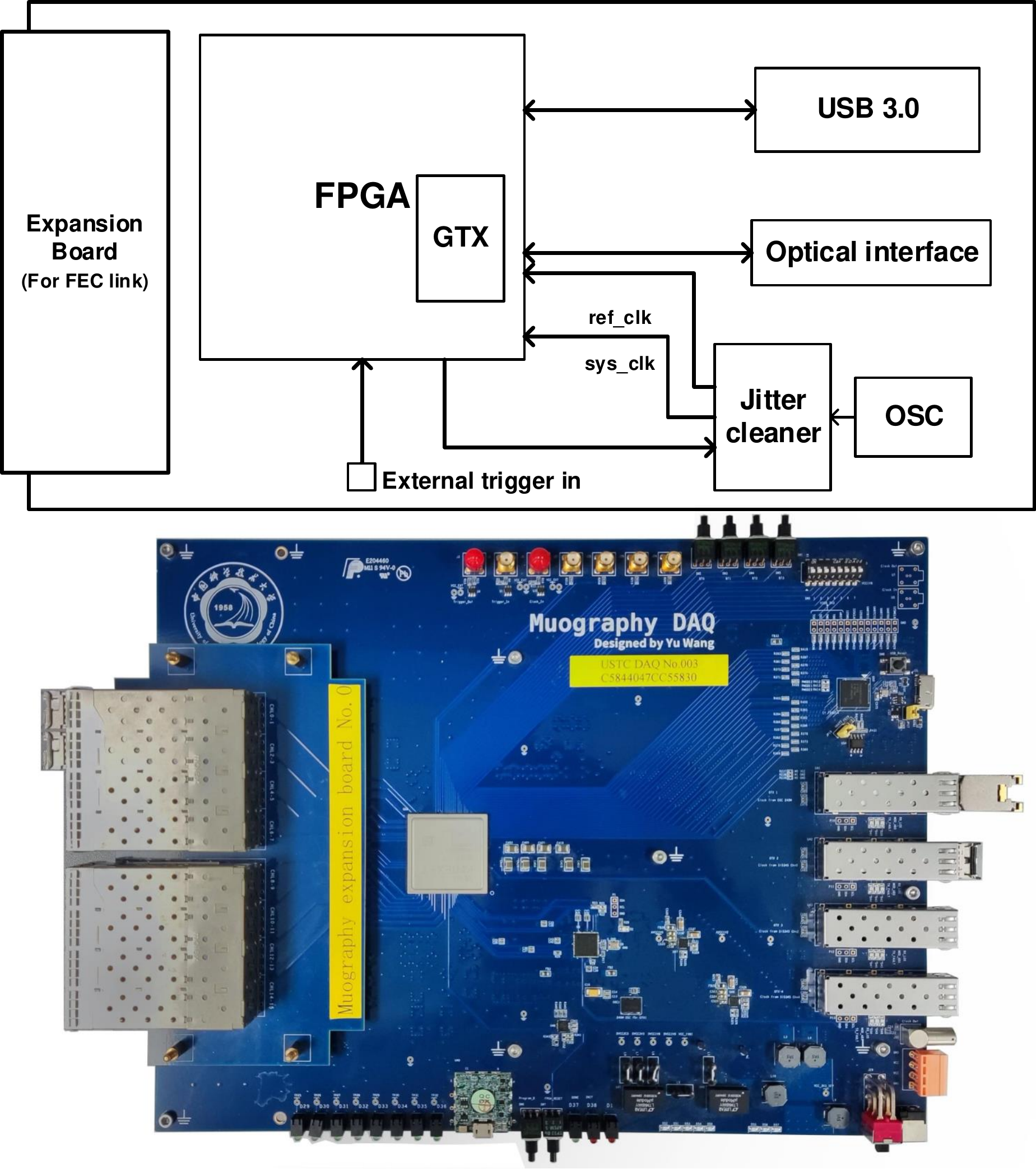}}
\caption{Schematic diagram and photograph of the DAQ board.}
\label{fig_7}
\end{figure}

The DAQ board was designed to distribute triggers and clocks to FECs, gather the data from different FECs, and pre-process the collected data. It was divided into two parts, namely, the expansion board designed to construct the front-end link part and the mainboard designed for control and communication. The link between DAQ and FEC was constructed using an optical fiber with a user-defined protocol. \figurename~\ref{fig_7} shows the structure of the DAQ board, which could support at most 16 FECs readout for a large-size tomography facility. The function of the FPGA was to implement the control and data reconstruction logic for both the front-end interface and the communication interface. The interfaces between DAQ board and server were USB 3.0 and optical fiber, which were selected according to the experimental settings. 

The DAQ board communicated with FECs via serial links using a user-defined protocol. The clock and trigger were encoded into the serial transmission data, and the clock data recovery (CDR) chip on the FEC decoupled the command and clock from this link. The trigger signals were generated by a pair of scintillators with PMTs (Photo-electron Multiply Tubes) and then fanned out to all FECs synchronously. When receiving data from the FECs, the DAQ board verified the frame structure and the checksum. A preset shaping time was used to select the region-of-interest data, and only one part of the SCA data was preserved and transmitted to the PC with the universal trigger ID, timestamp, and total checksum.

\subsection{Tomography Prototype System} 
\label{sub:tomography_prototype_system}
\begin{figure}[htbp]
\centerline{\includegraphics[width=3.5in]{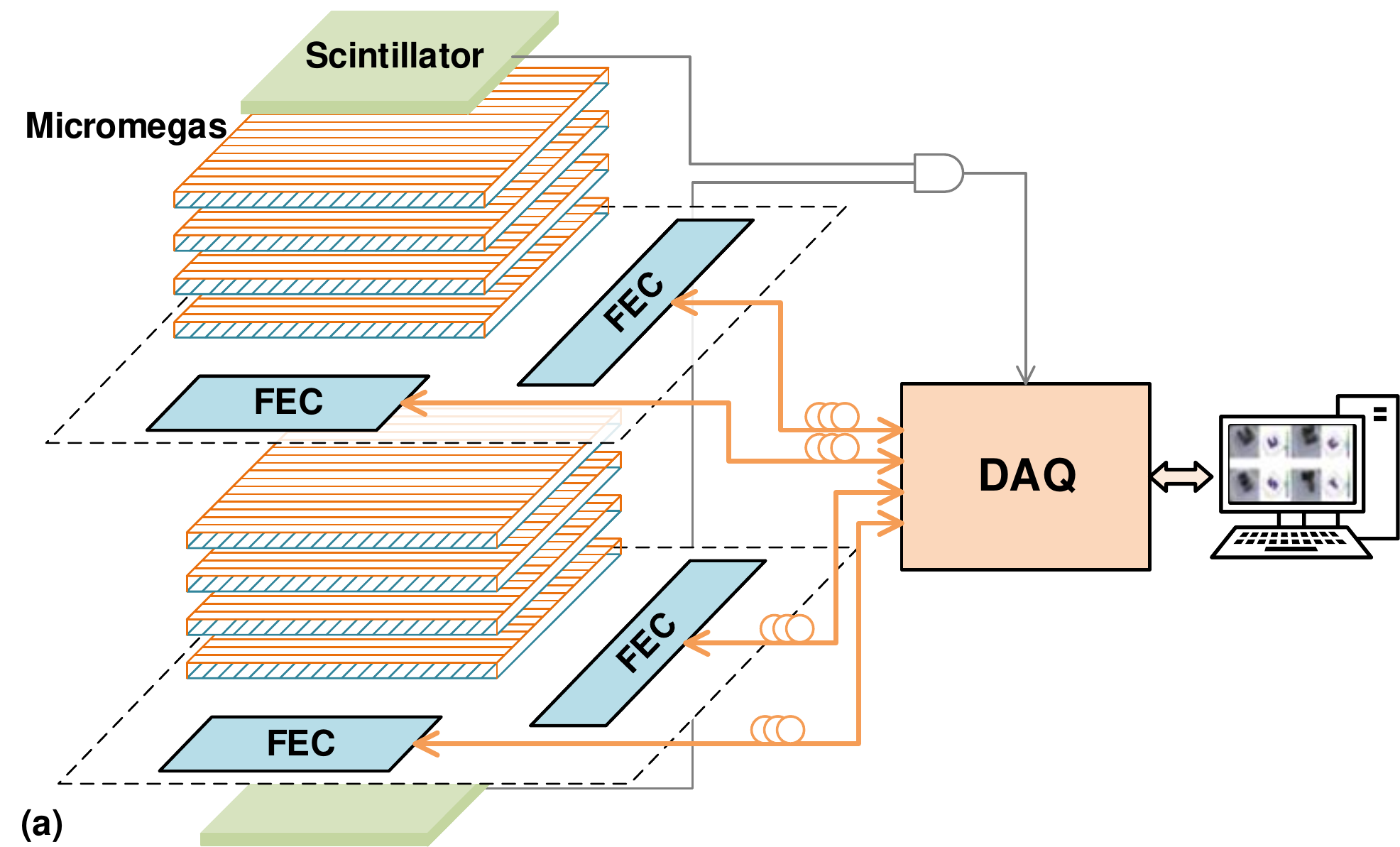}}
\centerline{\includegraphics[width=3.5in]{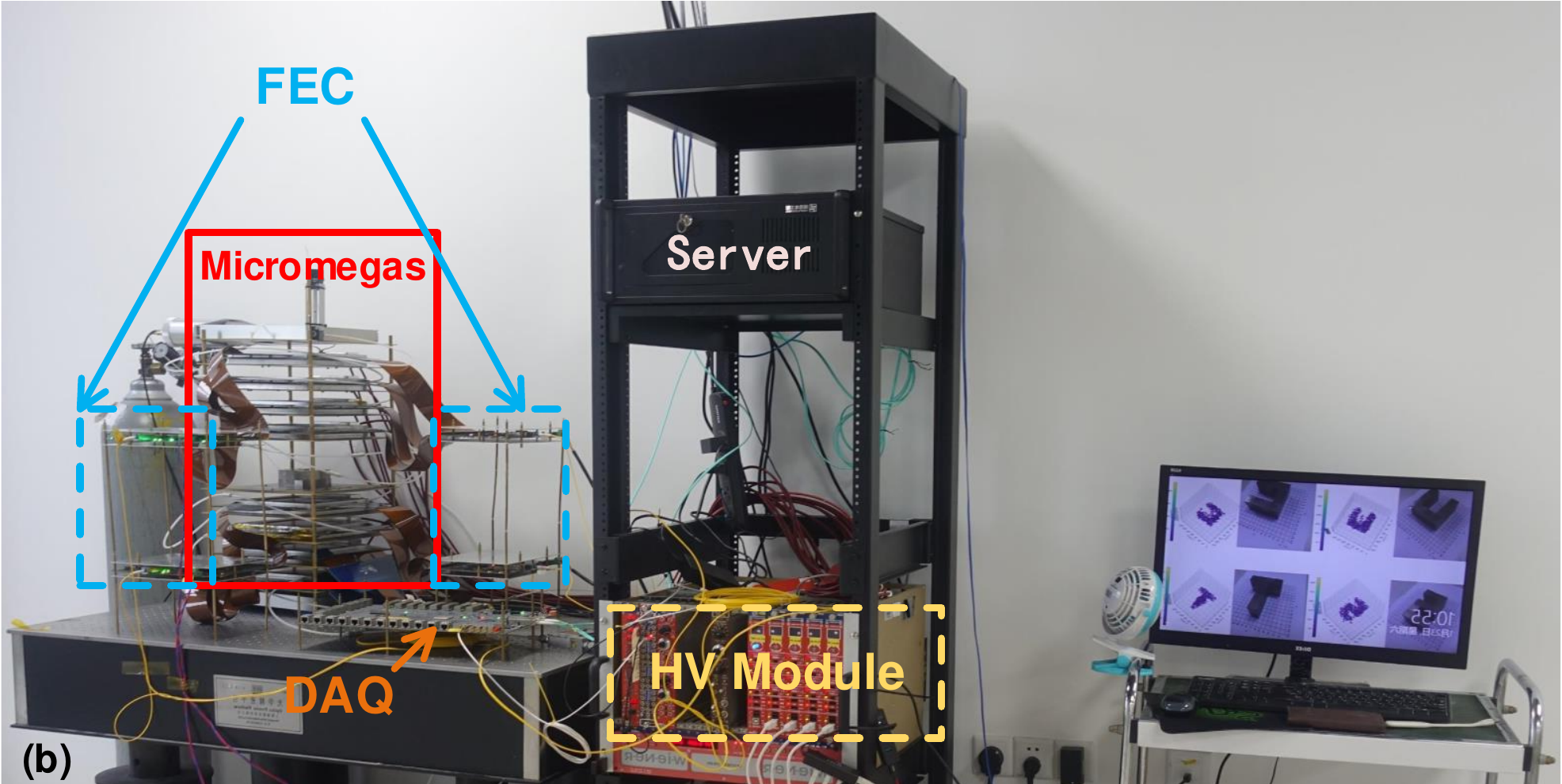}}
\caption{(a) Structure of the muon tomography prototype; (b) photograph of the prototype. The constructed system included Micromegas detectors, FECs, DAQ, server, and HV modules.}
\label{fig_8}
\end{figure}

A muon tomography prototype was constructed and named $\mathrm{\mu STC}$ (muon Scattering and Transmission imaging faCility). As shown in \figurename~\ref{fig_8}(a), this prototype contained eight Micromegas detectors, four FECs, a DAQ board, a server, two scintillators for trigger generation, and several high-voltage (HV) modules. When a muon passed through the active detector area, ionization signals were collected and sampled by the FEC and then transmitted to the DAQ board if a valid trigger was generated by the top and bottom scintillator planes. The photo of the prototype is displayed in \figurename~\ref{fig_8}(b).

\section{Experimental Results} 
\label{sec:experiments}

\subsection{Performance of the Readout Electronics} 
\label{sub:performance_of_the_readout_electronics}
\begin{figure}[htbp]
\centerline{\includegraphics[width=3.5in]{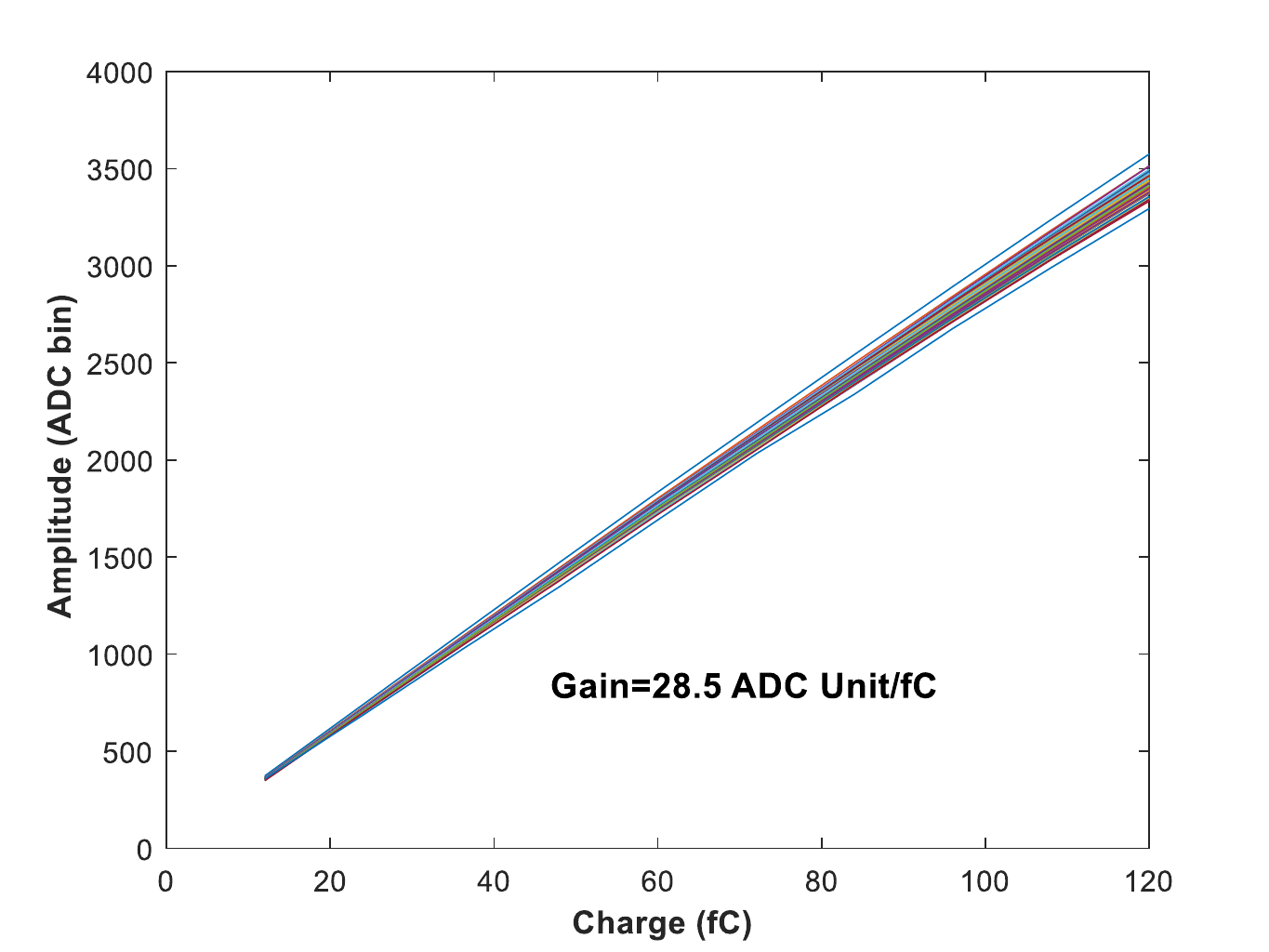}}
\caption{Calibration curve of a single AGET chip.}
\label{fig_9}
\end{figure}

\begin{figure}[htbp]
\centerline{\includegraphics[width=3.5in]{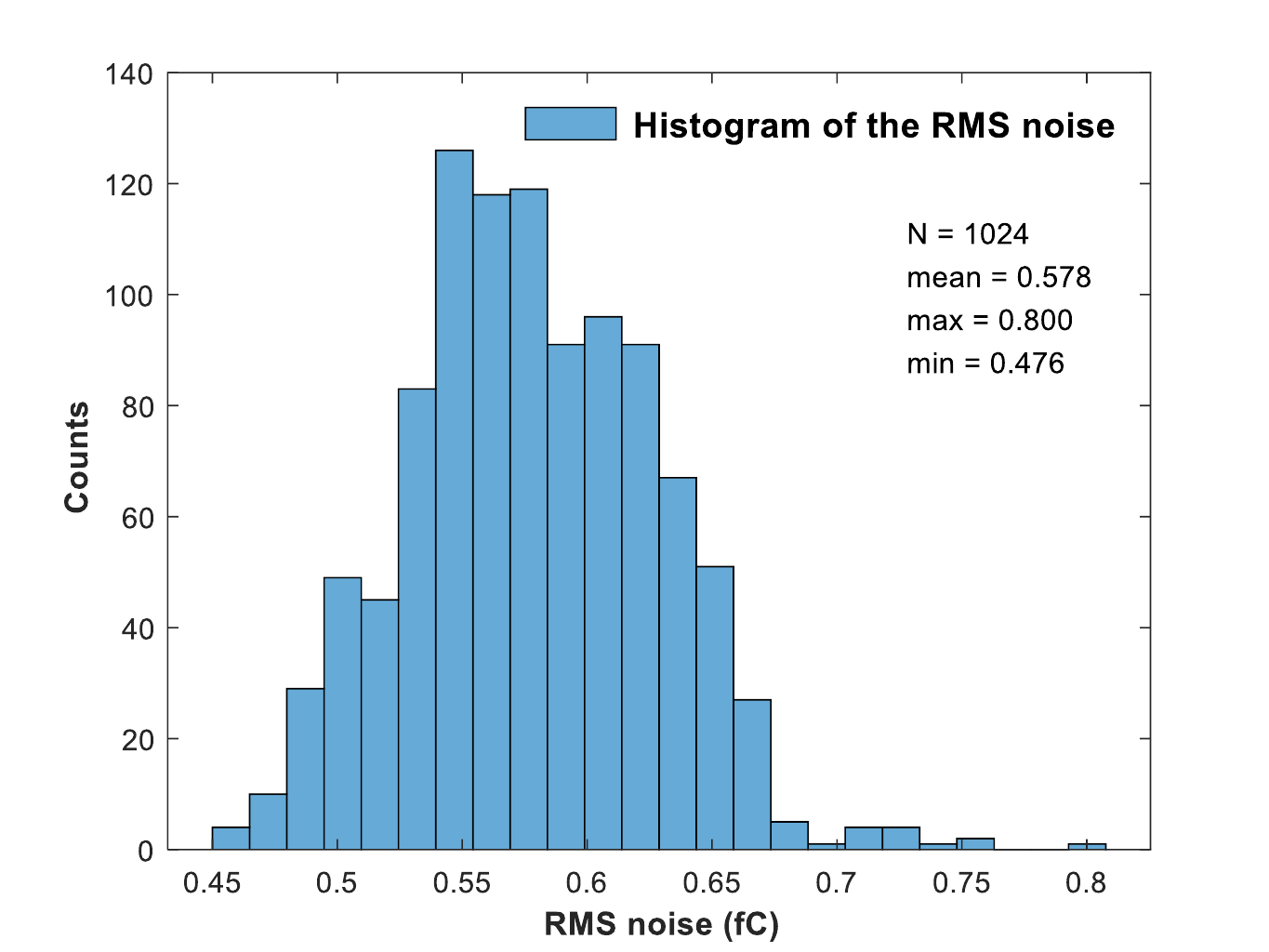}}
\caption{Distribution of the RMS noise of a single FEC connected to Micromegas detector.}
\label{fig_10}
\end{figure}

Each FEC was calibrated using an external calibration charge pulser. The gain and linearity performance of 64 channels from an FEC are presented in \figurename~\ref{fig_9}. The RMS of equivalent noise charge was measured when Micromegas detectors and FECs were connected, and as shown in \figurename~\ref{fig_10}, the RMS noise was less than 0.8 fC.

A pseudo-random binary sequence (PRBS) test was performed using the user-defined protocol between the FECs and DAQ board. A total of $3\times 10^{14}$ bits were transmitted without any error detected. The bit error ratio was lower than $10^{-14}$ with a confidence level of 95\%; thus, the stability of the transmission link could satisfy the requirement of long-time data acquisition.


\subsection{Decoding Test and Energy Spectrum} 
\label{sub:decoding_test_and_energy_spectrum}
\begin{figure}[htbp]
\centerline{\includegraphics[width=3.5in]{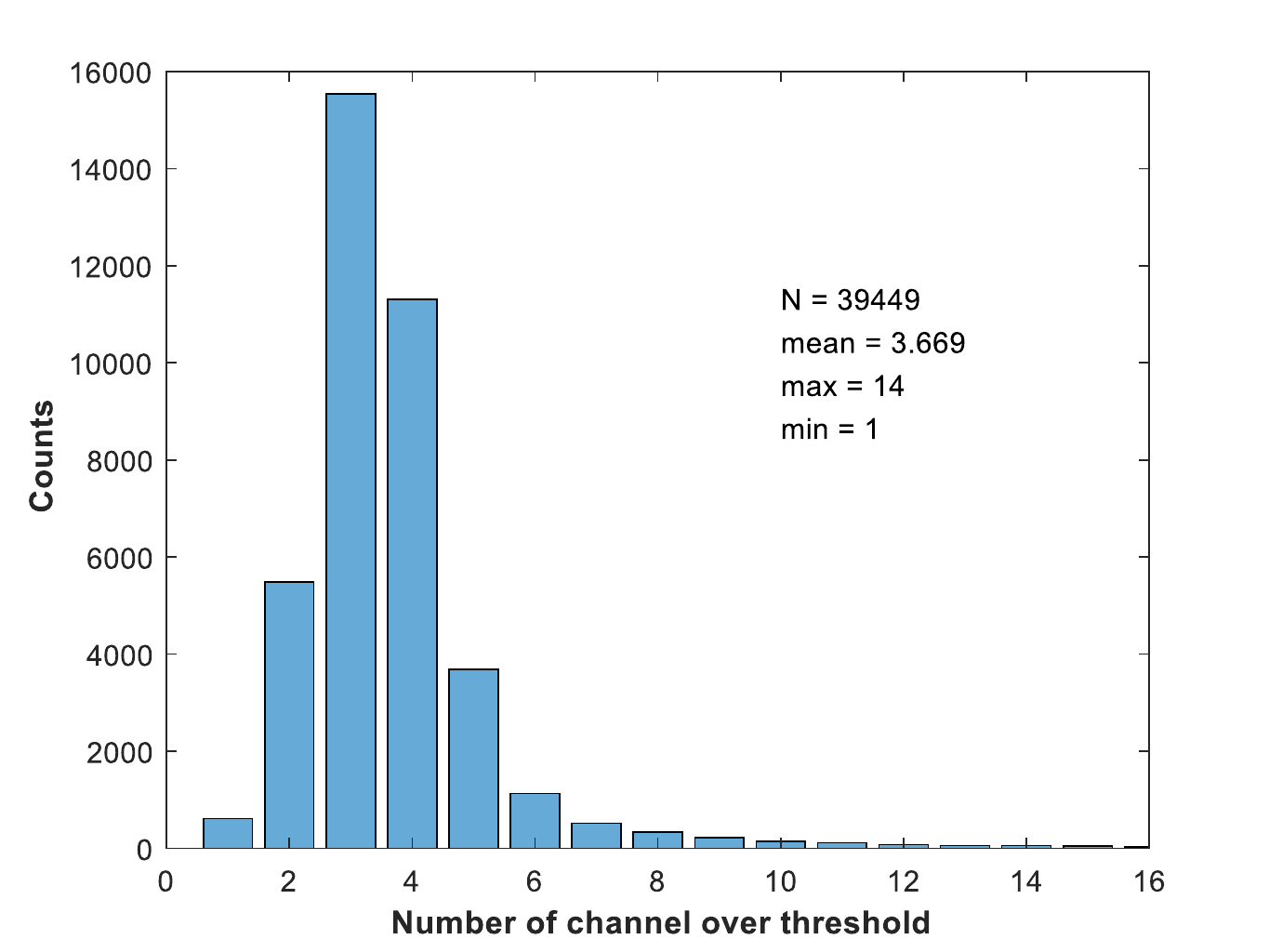}}
\caption{Distribution of the number of channels over thresholds in each event. The number of events that only hit one strip is 615.}
\label{fig_11}
\end{figure}

Since the detector channels were multiplexed, the hit strips could be decoded only if there were more than one channel with signal over the readout threshold. \figurename~\ref{fig_11} shows the distribution of hit strips, which implies that the decoding ratio was 98.4\%. This result indicates that the position encoding method can achieve almost lossless channel compression. 

\begin{figure}[htbp]
\centerline{\includegraphics[width=3.5in]{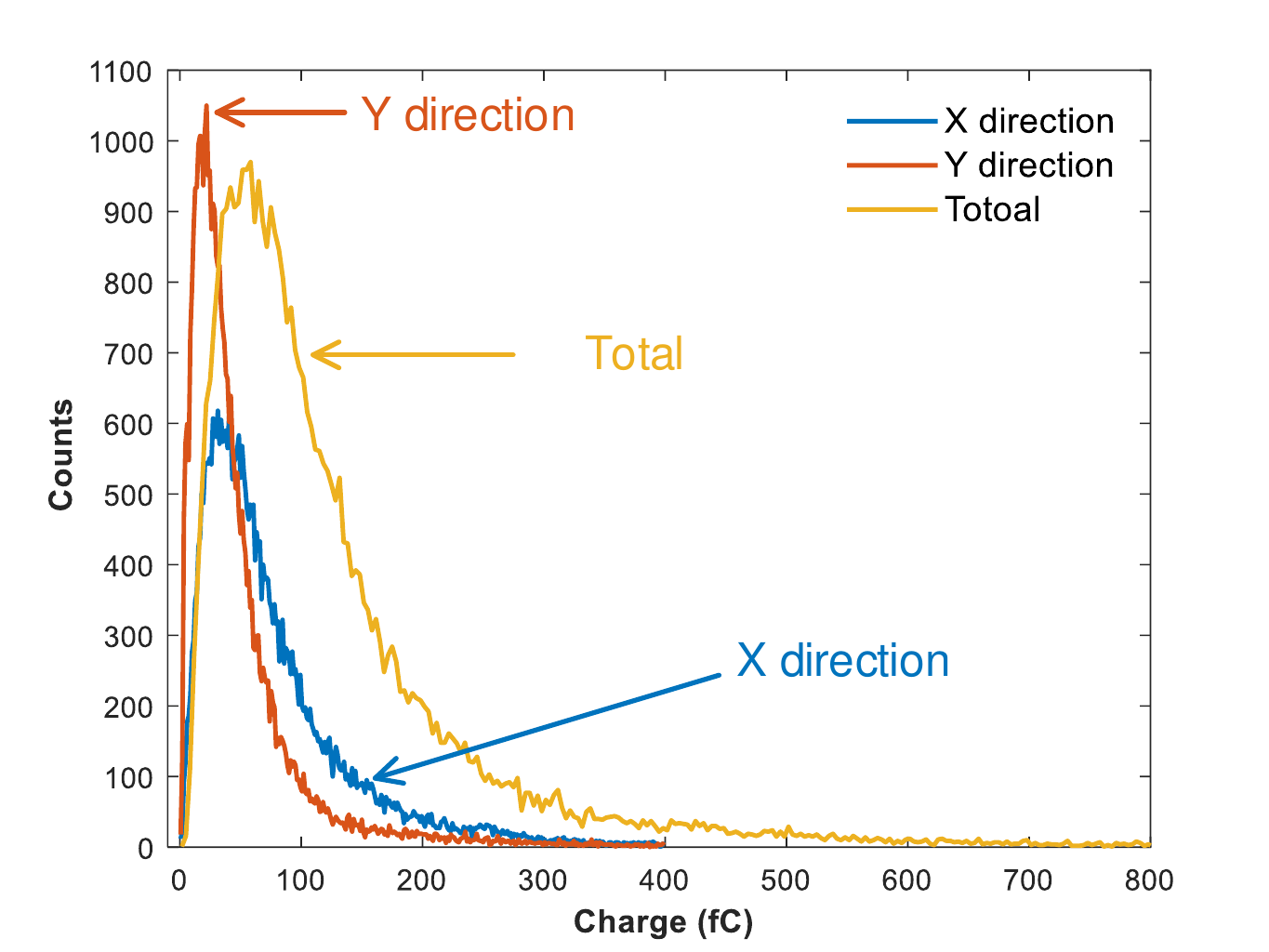}}
\caption{The energy spectrum of the cosmic ray. The total energy spectrum is the sum of the x-direction and y-direction of each event when both directions have signals over thresholds.}
\label{fig_12}
\end{figure}

Using the charge and decoded position information, the energy spectrum of cosmic-ray muons was reconstructed, as shown in \figurename~\ref{fig_12}. The Landau distribution shape indicates that the energy was reconstructed correctly by the position encoding method. The most probable value in the y-direction was less than that in the x-direction because the readout strips of the y-axis were below those of the x-axis, and the induced charge was small.

\subsection{Detector Alignment and Spatial Resolution} 
\label{sub:spatial_resolution}
\begin{figure}[htbp]
\centerline{\includegraphics[width=3.5in]{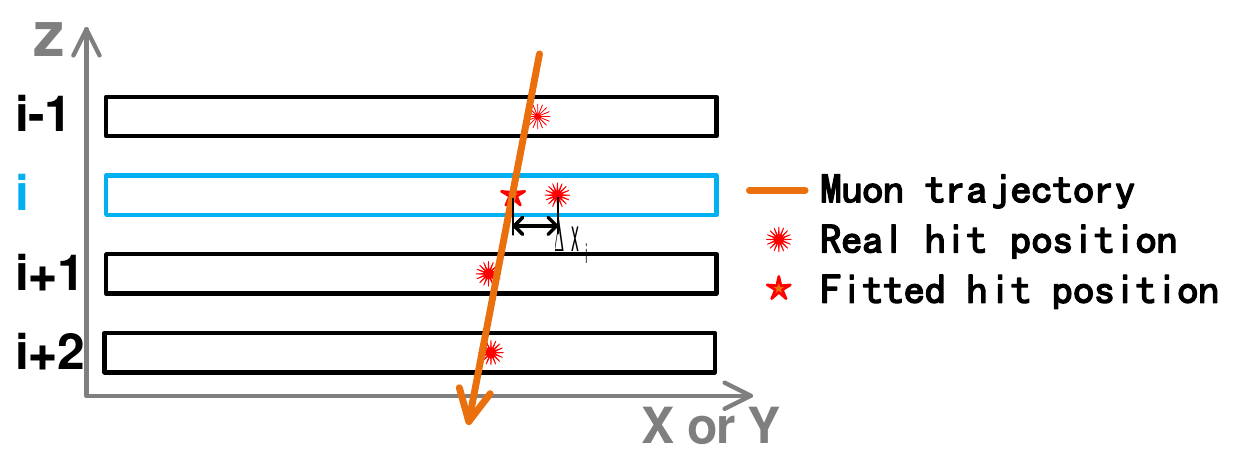}}
\caption{Illustration diagram of the spatial resolution calculation.}
\label{fig_13}
\end{figure}

A detector alignment algorithm was applied to correct the small offset and rotation caused by mechanical installation. In this algorithm, a difference between the real hit and recorded positions is divided into nonrandom and random parts, which correspond to the installation error and the spatial resolution, respectively. The purpose of the alignment is to minimize the nonrandom part. As shown in \figurename~\ref{fig_13}, let the $i^{th}$ layer be the target layer and fit the trajectory except for layer $i$. Then, the fitting result can be expressed as $x_{i_{f i t}}=k_{x, i} \times z_{i}+b_{x, i}$, where $k_{x, i}$ is defined by (\ref{ref_eq_kxi}), and $b_{x, i}$ is calculated as $b_{x,i}=\bar{x}-k_{i} \cdot \bar{z}$. Accordingly, the offset was defined as $\Delta x_{i}=x_{i_ {f i t}}-x_{i_{ h i t}}$, and a good alignment parameter should minimize the offset among all events. A gradient descent method was applied to the alignment, and the loss function was defined by (\ref{eq_gd_loss}).

\begin{equation}
k_{x, i}=\frac{\sum_{j \neq i}^{n}\left(z_{j}-\bar{z}\right) x_{j_{h i t}}}{\sum_{j \neq i}^{n}\left(z_{j}-\bar{z}\right)^{2}}
\label{ref_eq_kxi}
\end{equation}

\begin{equation}
\begin{aligned}
  L = \sum_{event = 1}^M \left[\sum_{i=0}^7\left( \Delta x_{i} ^2 + \Delta y_{i}^2 \right)\right]
\end{aligned}
  \label{eq_gd_loss}
\end{equation}

\begin{figure}[htbp]
\centerline{\includegraphics[width=3.5in]{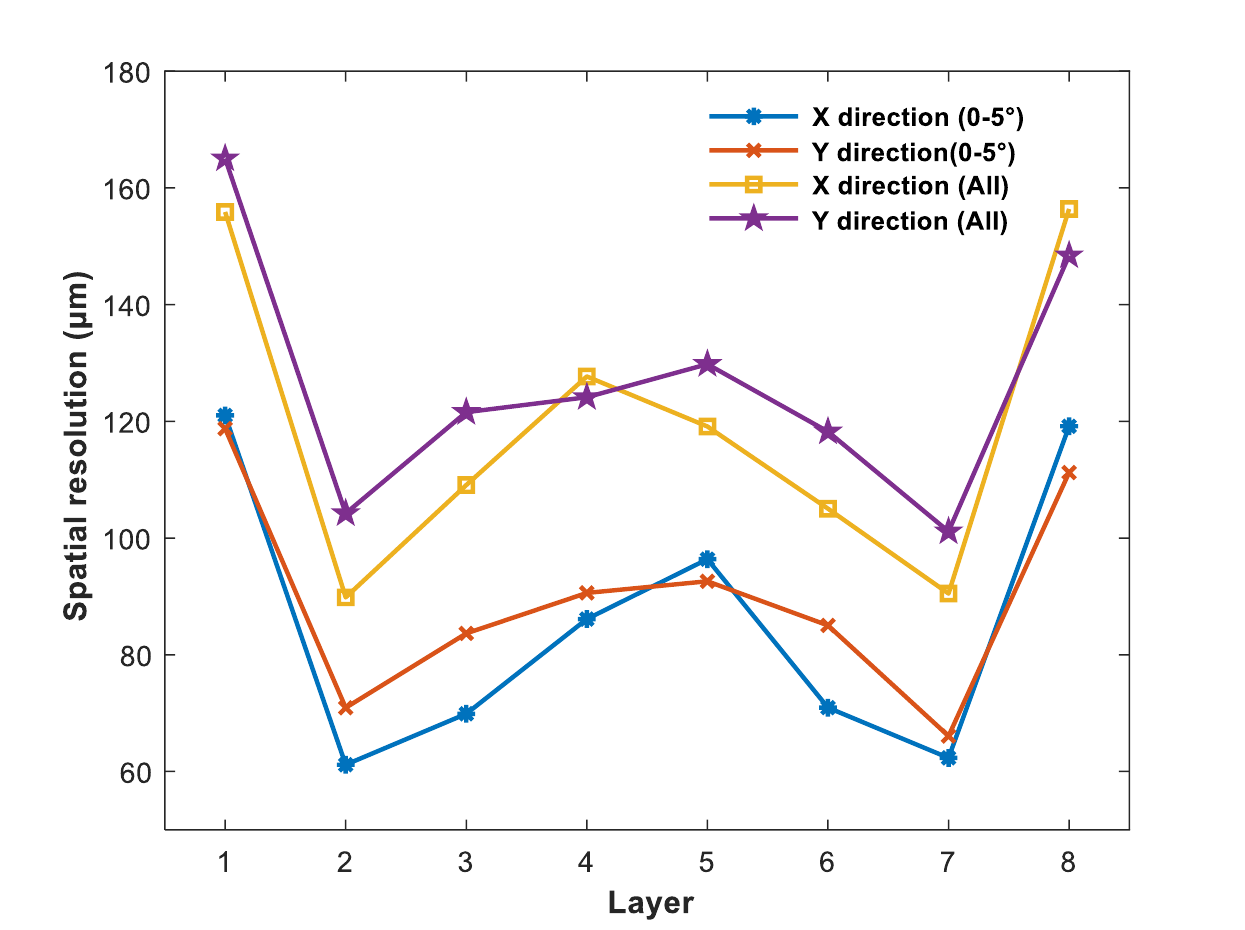}}
\caption{Spatial resolution with the incident angle in the range of 0–5 degree and all.}
\label{fig_14}
\end{figure}

After alignment, the spatial resolution was defined as a standard deviation of $\Delta x_i$ ($\sigma(\Delta x_i)$) and divided into two parts: the deviation of trajectory fitting ($\sigma (x_{fit})$) and the residual of the detector under test ($\sigma(x_{hit})$). As mentioned before, $x_{i_{ f i t}}$ was independent of $x_{i_{hit}}$, so the resolution was calculated by $\sigma^{2}(\Delta x_i) = \sigma^{2}(x_{i_{fit}}) + \sigma^{2}(x_{i_{ hit}}) = \sum_{j=1}^{n} B_{ij}*\sigma^{2}(x_{i_{ hit}})$, where $B_{i j}$ is defined by (\ref{eq_bij}), and $\sigma^{2}(\Delta x_{i})$ is calculated from the measured data; the spatial resolution of the detector was calculated by solving linear equations. \figurename~\ref{fig_14} shows the spatial resolution of the proposed muon tomography system with an incident angle of between zero to five degrees and all angles. Under a small incident angle, the mean spatial resolution was about $\mathrm{85 ~\mu m}$, which agreed with the previous test with electron beams~\cite{bib:bib14}.
\begin{equation}
B_{i j}=\left\{\begin{array}{c}
1, i=j \\
{\left[\left(z_{i}-\bar{z}\right) a_{i j}+\frac{1}{n-1}\right]^{2}, i \neq j}
\end{array}\right.
 \label{eq_bij}
\end{equation}


\subsection{Tomography experiments} 
\label{sub:tomography_experiments}
\begin{figure}[htbp]
\centerline{\includegraphics[width=3.5in]{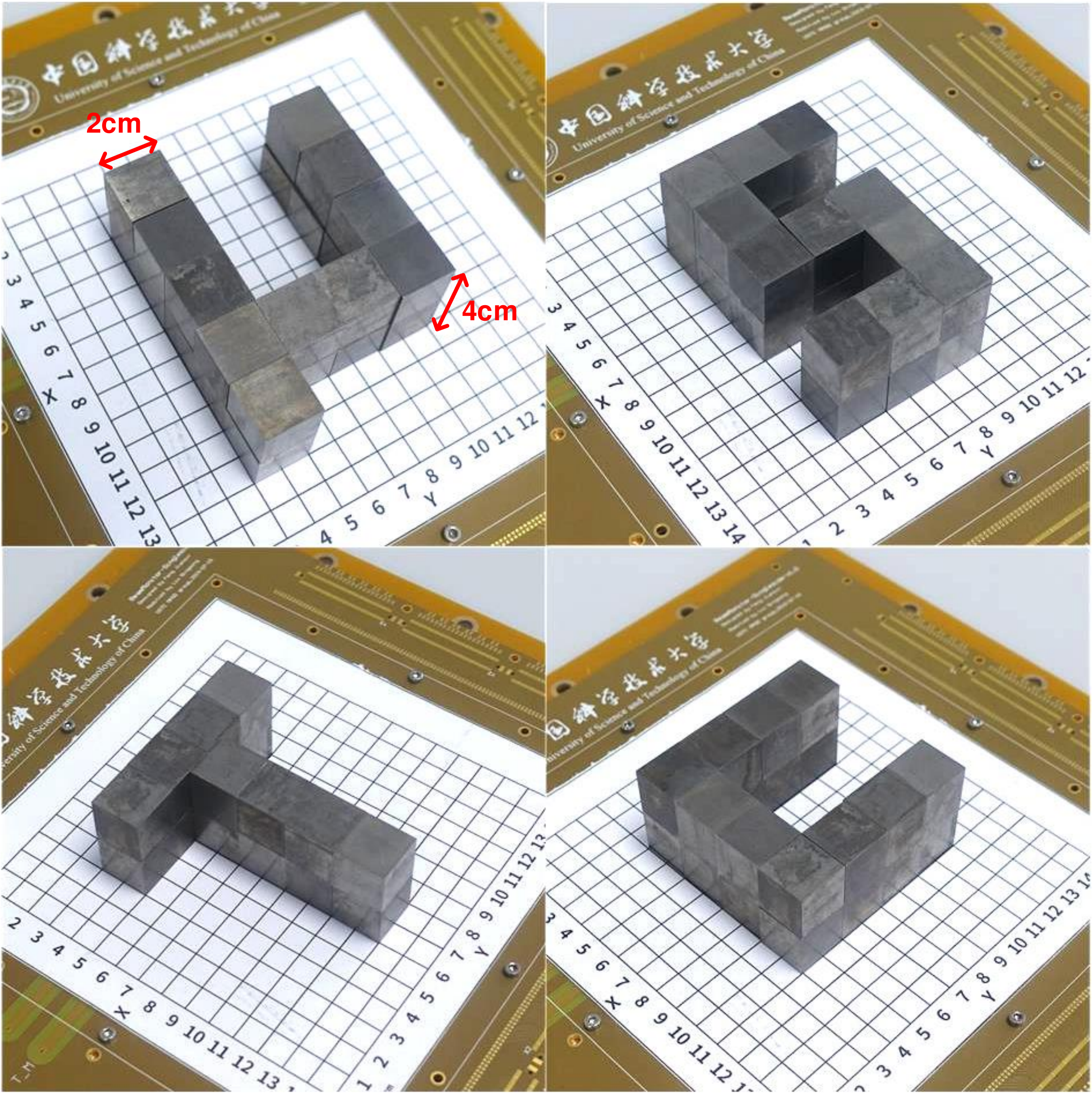}}
\caption{Photograph of four groups of samples under test.}
\label{fig_15}
\end{figure}

As shown in \figurename~\ref{fig_15}, four groups of samples were imaged with the proposed prototype.  These samples were made of small tungsten cube, and the size of each cube was $\mathrm{2~cm \times 2~cm \times 2~cm}$. In the vertical direction, the thickness of these samples was 4 cm. In this research, the imaging results were reconstructed by the PoCA algorithm \cite{bib:bib4}, which assumes that the Coulomb scattering has happened just once at the closest distance between the incident and scattering trajectories. The scattering angle of the event was calculated as an angle of two trajectories and used to judge whether the scattering had happened. Due to the single Coulomb scattering assumption, some of the PoCA points might be reconstructed improperly. To remove the incorrect PoCA points, a k-NN (k-Nearest Neighbors) like algorithm was applied for image reconstruction. A PoCA point was valid only when the number of nearby PoCA points exceeded a reasonable threshold. By using this method, the fake reconstruction points caused by the PoCA algorithm and the error hit position were rejected.

\begin{figure}[htbp]
\centerline{\includegraphics[width=3.0in]{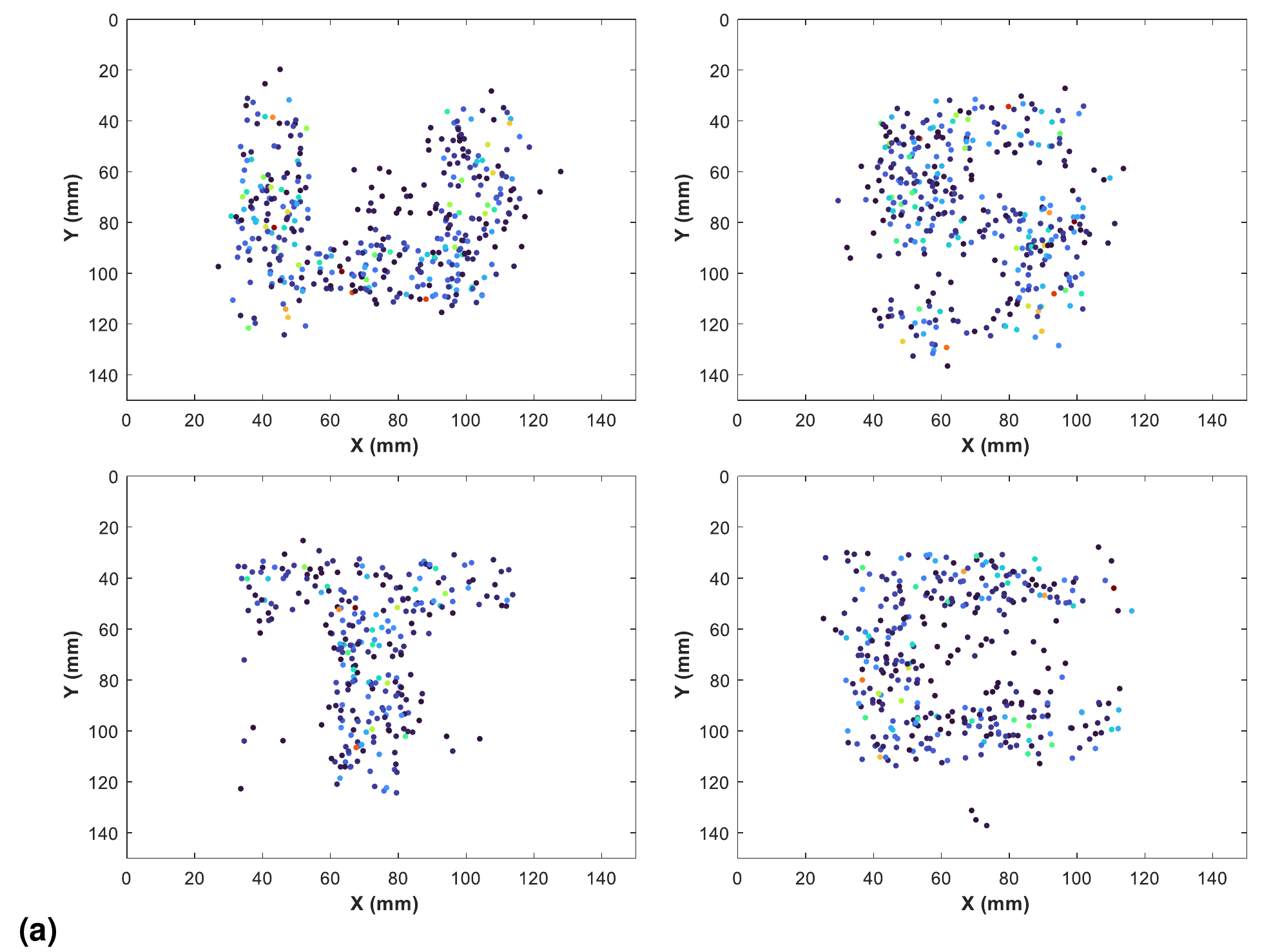}}
\centerline{\includegraphics[width=3.0in]{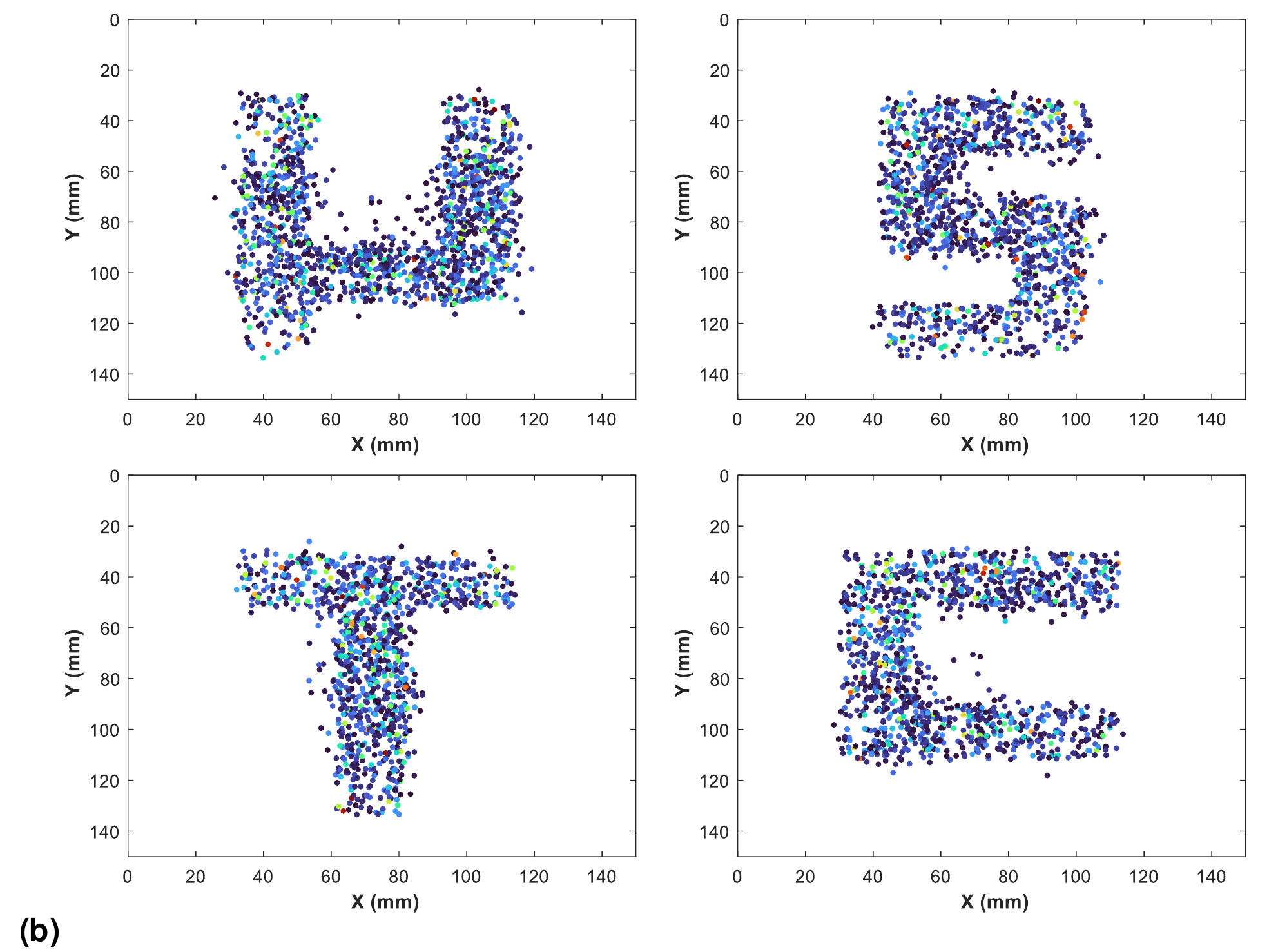}}
\caption{Tomography results after four-hour muon exposure; (b) result after 24-hour muon exposure.}
\label{fig_16}
\end{figure}

The reconstruction results of these samples of the four-hour test are shown in \figurename~\ref{fig_16}(a). The results indicated that under a reasonable threshold for the scatter angle cut, each sample could be appropriately reconstructed. The previous research has shown that the event rate of the edge area is only 10\% or less of the center area. Due to this effect, in \figurename~\ref{fig_16}(a), the bottom left corner of the $\mu $-shape is missing. When the test time was extended to 24 hours, the imaging results were as shown in \figurename~\ref{fig_16}(b), where it can be seen that each sample was reconstructed accurately.

\begin{figure}[htbp]
\centerline{\includegraphics[width=3.0in]{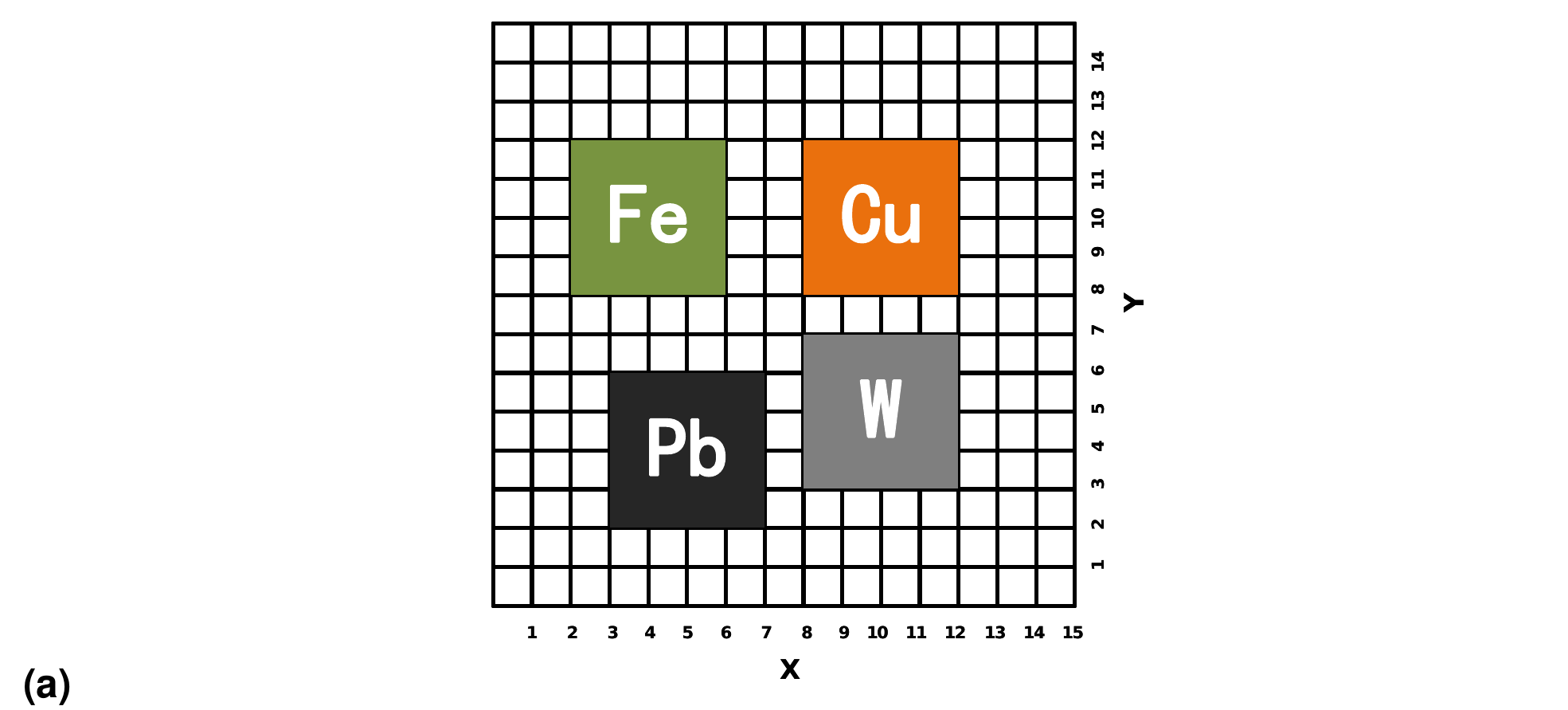}}
\centerline{\includegraphics[width=3.0in]{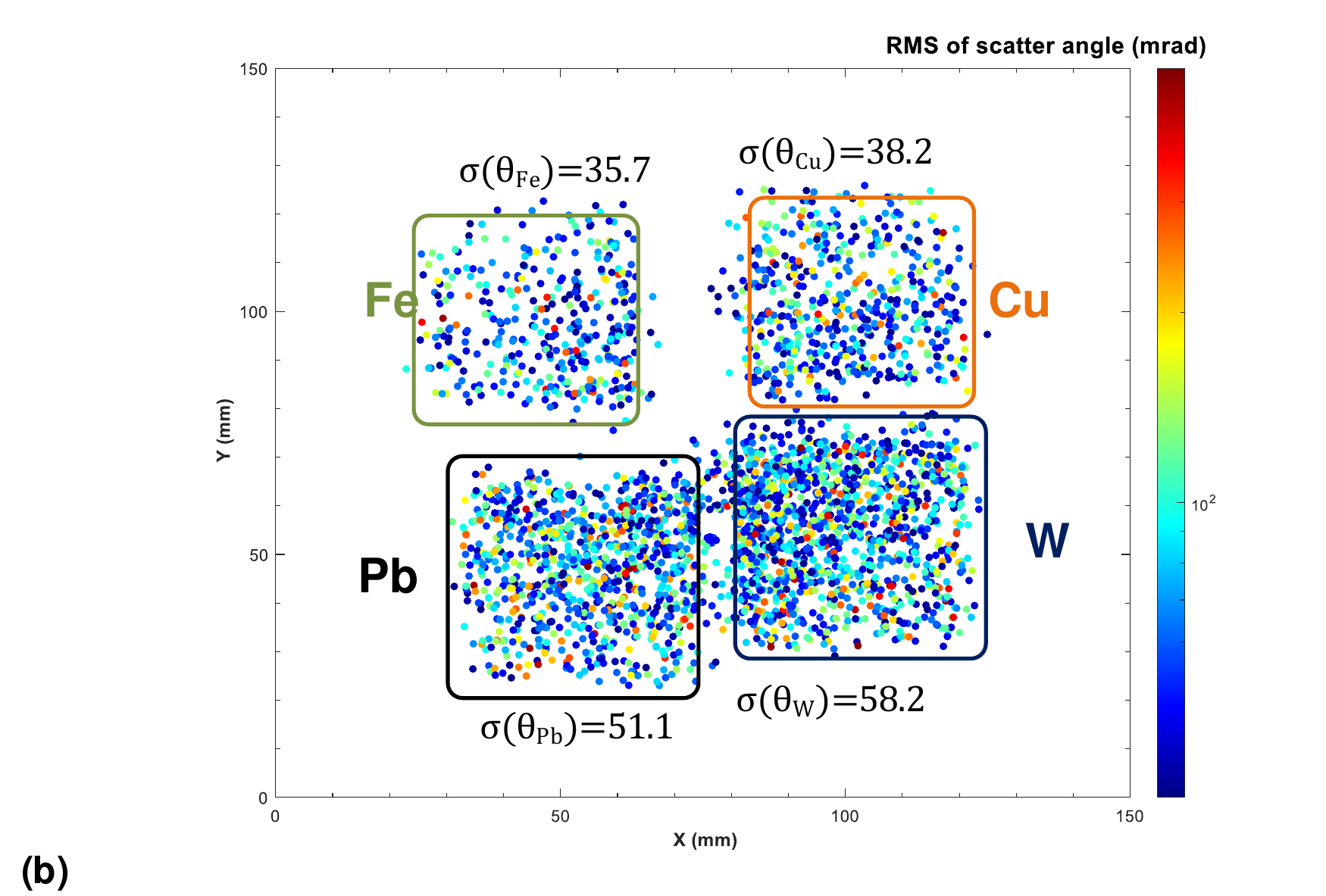}}
\caption{(a) Top view of the samples’ positions; (b) the reconstruction result and the standard deviation of the projected scattering angle of each sample.}
\label{fig_17}
\end{figure}

Another important application of muon tomography is to distinguish different materials. When thicknesses of different materials are equal, the RMS width of the scattering angle is inversely proportional to the square root of radiation length. Thus, the RMS width of the projected scatter angle can be viewed as a parameter to separate different materials. To verify the scatter angle reconstruction ability, four samples made of iron (Fe), copper (Cu), lead (Pb), and tungsten (W) were imaged, and the size of each sample was $\mathrm{4~cm \times 4~cm \times 4~cm}$, as shown in \figurename~\ref{fig_17}(a). The tomography result and the standard deviation of scattering angle in each area are shown in \figurename~\ref{fig_17}(b). The order of the radiation lengths was as follows $X_{F e}>X_{C u}>X_{P b}>X_{W}$; meanwhile, the RMS values of the scattering angle had an opposite order: $\sigma(F e)<\sigma(C u)<\sigma(P b)<\sigma(W)$.  In the reconstruction process, a threshold cut was set to eliminate the error point. Thus, some of the scattering points with small scattering angles were missing, and the numbers of PoCA points were different for these materials.


\section{Conclusion and Discussion} 
\label{sec:conclusion_and_discussion}
In this study, a high spatial resolution tomography prototype based on Micromegas detectors is designed and implemented. The long-term stability and fine resolution of the proposed design are verified for the thermal bonding Micromegas. In addition, a novel multiplexing method based on position encoding is developed to reduce a large number of readout channels. The results have indicated that the requirement for readout channels can be reduced by one order of magnitude without hit information loss by using the proposed method.  Moreover, a scalable readout system is designed and implemented. The RMS noise of the readout electronics is about 0.8 fC, and the dynamic range is up to 120 fC. The spatial resolution of the detectors with encoding readout is about one hundred micrometers. In addition, the proposed prototype is verified by imaging experiments. The results indicate that the proposed prototype can image and distinguish materials of objects with a size of several centimeters. 

The prototype presented in this paper indicates that high-resolution Micromegas detectors can be a promising solution for muon tomography, while the multiplexing readout method can reduce system complexity. The current research and developing effort has been ongoing toward a large muon tomography facility for application purposes. In future work, the micro-TPC algorithm will be introduced to improve the spatial resolution, and the self-trigger mode could be further improved to make the facility more compact.

\section*{Acknowledgment}
This work was partially performed at the University of Science and Technology of China (USTC) Center for Micro and Nanoscale Research and Fabrication, and the authors thank Yu Wei for his help in the nanofabrication steps for Ge coating, and Dianfa Zhou for his help on the laser cutting of thermal spacers. 

\ifCLASSOPTIONcaptionsoff
  \newpage
\fi

\end{document}